          [2001/04/25 1.1 (PWD)]
\documentclass[letterpaper,twocolumn]{esapub} 

\usepackage{natbib}
\usepackage{graphicx}

\title{An {\em INTEGRAL} Observation of the Black Hole Transient
4U 1630--47 and the Norma Region of the Galaxy}
\author{John A. Tomsick}
\author{Richard Lingenfelter}
\affil{Center for Astrophysics and Space Sciences, Code
0424, University of California at San Diego, La Jolla, CA,
92093, USA (e-mail: jtomsick@ucsd.edu)}
\author{Stephane Corbel}
\affil{Universit\'e Paris VII and Service d'Astrophysique,
CEA Saclay, 91191 Gif sur Yvette, France}
\author{Andrea Goldwurm}
\affil{Service d'Astrophysique, CEA Saclay, 91191 Gif 
sur Yvette, France}
\author{Philip Kaaret}
\affil{Harvard-Smithsonian Center for Astrophysics, 60 Garden Street,
 Cambridge, MA, 02138, USA}

\begin{document}

\keywords{accretion, accretion disks --- black hole physics: general ---
stars: individual (4U 1630--47) --- stars: black holes --- X-rays: stars}

\maketitle

\begin{abstract}

We report on results from a 2003 February {\em INTEGRAL} observation 
of the black hole candidate (BHC) 4U 1630--47 and the high energy
sources in the Norma Region of the Galaxy and on {\em RXTE} and radio 
observations of 4U 1630--47.  To date, there have been 38 new or 
previously poorly studied sources found by {\em INTEGRAL} (the 
``IGR'' sources), and 15 of the IGR sources are in the field of view 
of our observation.  We detect 12 of the 15 sources, and we speculate 
that a subset of the IGR sources may be part of a new population of 
persistent, hard, intermediate luminosity X-ray sources.  During our 
observations, 4U 1630--47 was in an intermediate state with strong 
soft and hard components, a power-law index of $\Gamma = 2.11$-2.18, 
a 4\% rms level of timing noise (0.1-10 Hz), and the source was
not detected in the radio band.  We discuss the evolution of the 
energy spectrum during the 293 ks {\em INTEGRAL}
observation, and we report on rapid, high amplitude flaring 
behavior that has been seen for much of the 2002-2004 outburst.

\end{abstract}

\section{Introduction}

The X-ray flux observed from transient X-ray binaries changes by 
several orders of magnitude between quiescence and outburst, providing 
an opportunity to learn about black hole candidates (BHCs), accretion 
onto compact objects, and the processes responsible for high energy 
emission.  Although these objects are known to emit hard X-rays 
and gamma-rays, the origin of this emission is not well-understood.  
The high energy emission provides a probe of the regions of space 
close to the compact object, allowing us to study strong gravity 
and jet formation processes.  4U 1630--47 is among the most active 
of the BHC X-ray transients and has produced strong hard X-ray 
emission during its 17 detected outbursts 
\citep{tk00,oosterbroek98,kuulkers97,parmar97} as well as highly 
polarized radio emission \citep{hjellming99}, indicating the presence 
of jets.  The current outburst from 4U 1630--47, which began in 2002 
September, is one of the brightest and longest recorded outbursts 
from this system, and we had the opportunity to observe 4U 1630--47 
with {\em INTEGRAL} in 2003 February.  

Due to the large {\em INTEGRAL} field of view, we also detect 
many other hard X-ray sources in the vicinity of 4U 1630--47.
Several of the sources were unknown or poorly studied prior to 
{\em INTEGRAL} and have ``IGR'' designations.  Although we are 
just beginning to understand the nature of the IGR sources, many 
of these appear to be highly absorbed ($N_{\rm H} = 10^{23-24}$
cm$^{-2}$), and some of them are bright IR sources
\citep[e.g.,][and papers by Foschini, Walter, and others in
these proceedings]{mg03,walter04,patel04}.  The concentration of 
IGR sources near 4U 1630--47 is almost certainly due to the 
location of the source in the Galactic plane ($l = 336.91^{\circ}$, 
$b = +0.25^{\circ}$).  This is in the Norma region of the Galaxy,
and the line of sight to 4U 1630--47 is tangent to the Norma-Scutum 
Galactic arm.  Below, we present our results on:  The hard 
X-ray sources detected by {\em INTEGRAL} in the Norma region; and 
X-ray and radio observations of 4U 1630--47.

\section{Observations}

We obtained the following observations of 4U 1630--47 for this 
project:\\
\noindent
$\bullet$~{\em INTEGRAL}:  Our observations with {\em INTEGRAL} 
lasted from UT 2003 February 1, 5.7 h to UT 2003 February 5, 
7.9 h, with a total of 293 ks on source.\\
\noindent
$\bullet$~{\em RXTE}: 4U 1630--47 has been monitored throughout 
the current outburst with the {\em Rossi X-ray Timing Explorer 
(RXTE)}.  Five {\em RXTE} observations occurred during or 
within a day of the {\em INTEGRAL} observation, yielding 43 ks 
with the Proportional Counter Array (PCA) and 28 ks with the 
High Energy X-ray Timing Experiment (HEXTE).\\
\noindent
$\bullet$~{Radio}:  We observed 4U 1630--47 on 2003 January 26 
and 27 at 4.8 and 8.6 GHz.  The target was not detected with 
an rms level of 0.1 mJy.

\section{Results}

\subsection{The Norma Region Sources}

\begin{figure}[ht]
\centerline{\includegraphics[width=1.0\linewidth,angle=0]{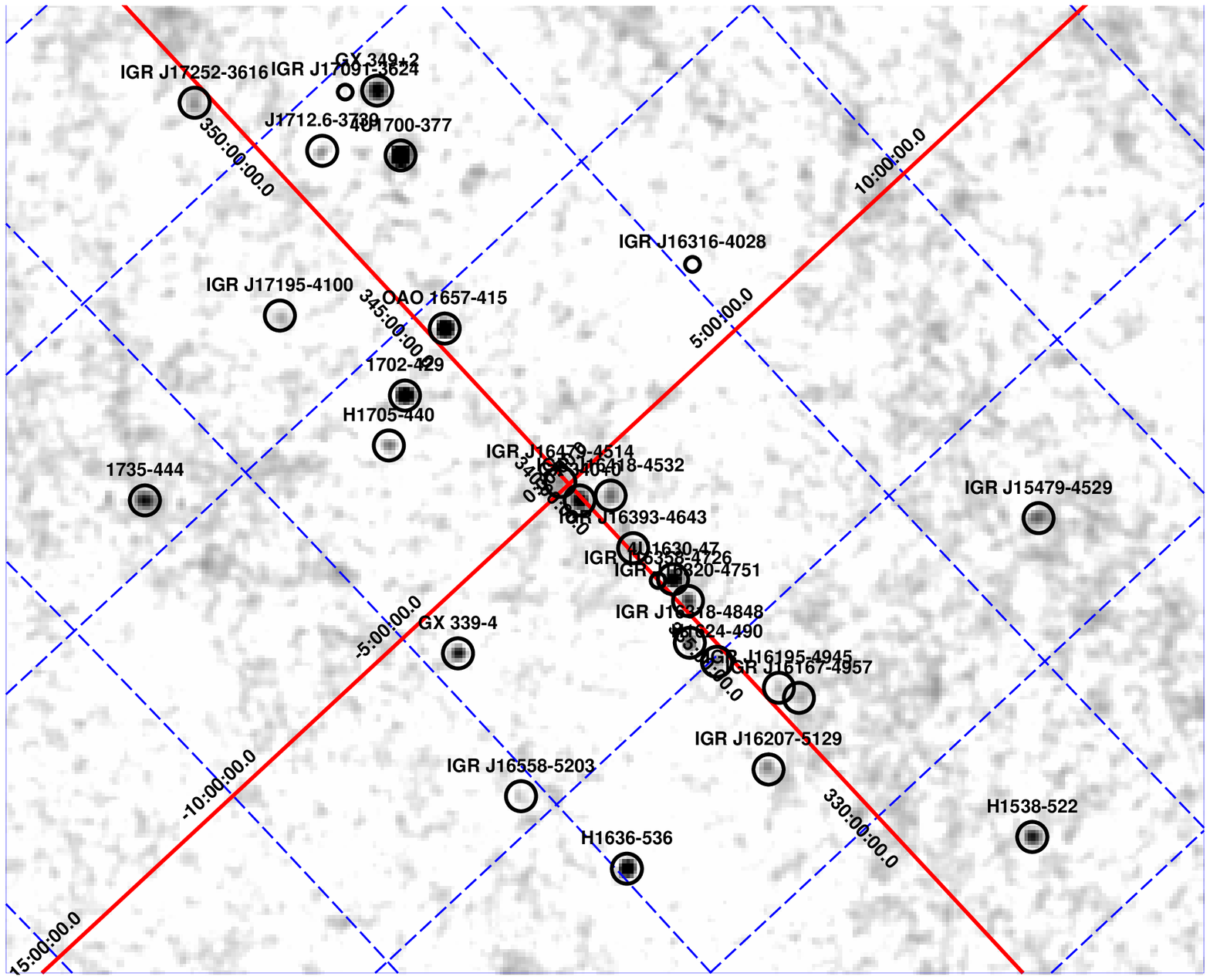}}
\caption{20-40 keV ISGRI significance image from our 2003
February 293 ks INTEGRAL observation.  Detected sources that we 
identified are marked with larger circles, and IGR sources reported 
in other observations that are not detected in our observation are 
marked with smaller circles. 
\label{fig:image_full}}
\end{figure}

\begin{figure}[ht]
\centerline{\includegraphics[width=1.0\linewidth,angle=0]{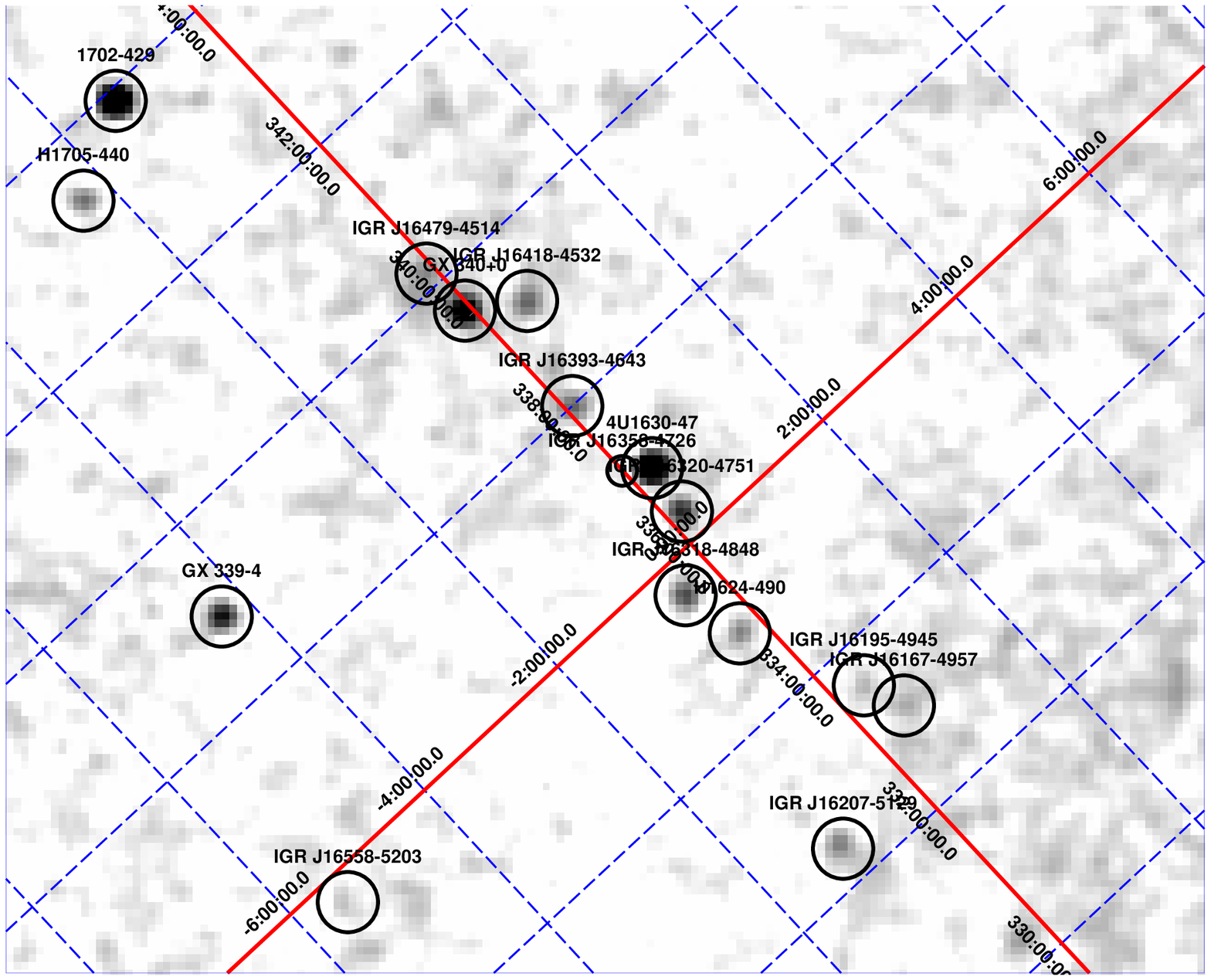}}
\caption{Close-up of the Norma Region.  The image intensity
scale and source markings are the same as for 
Figure~\ref{fig:image_full}, and this is also a 20-40 keV image.
\label{fig:image_zoom}}
\end{figure}

Figure~\ref{fig:image_full} shows the 20-40 keV IBIS image from 
our 2003 February observation of 4U 1630--47.  The field of view 
covered by IBIS includes 15 of the 38 IGR sources that have 
been found to date (as of 2004 March).  Thus, this relatively 
long single exposure provides a useful study of a significant 
subset of the IGR sources.  Table~1 lists the 15 IGR sources 
along with detection significances (SNR = signal-to-noise
ratio) and ISGRI count rates or upper limits.  We detect 12 
out of 15 of the IGR sources, and we note that we used these 
data to report the discovery of three of the sources 
\citep{tomsick03_iauc,tomsick04}.  In addition, the table gives 
information about identifications or possible identifications
with other high energy sources or with sources at other
wavelengths.  In many cases, this information is
in the discovery references (also given in the table), and 
we also searched the SIMBAD database to look for likely or 
possible identifications.  The density of sources within 
about $4^{\circ}$ of 4U 1630--47 is especially high, and 
Figure~\ref{fig:image_zoom} shows a close-up of this region.

\begin{table*}[ht]
\begin{center}
\caption{4U 1630--47 and IGR Sources}
\footnotesize
\begin{tabular}{lcccc} \hline \hline
Name             & SNR    & ISGRI Rate$^{\mathrm{a}}$ & Identification or Comment & Reference\\ \hline
4U 1630--47      & 123.0  & 3.75   & 4U 1630--47 & --\\ \hline
IGR J15479--4529 & 17.4   & 0.69   & 1RXS J154814.5--452845 & \cite{tomsick04}\\
IGR J16167--4957 & 6.9    & 0.24   & -- & \cite{walter04}\\
IGR J16195--4945 & 5.8    & 0.20   & AX J1619.4--4945, BIa star? & \cite{walter04}\\
IGR J16207--5129 & 11.5   & 0.42   & A1IVe star? & \cite{walter04}\\
IGR J16316--4028 & $<$4   & $<$0.2 & 3EG 1631--4033 (?) & \cite{rg03}\\
IGR J16318--4848 & 24.4   & 0.77   & sgB[e] star$^{\mathrm{b}}$ & \cite{courvoisier03}\\
IGR J16320--4751 & 32.7   & 1.01   & AX J1631.9--4752 & \cite{tomsick03_iauc}\\
IGR J16358--4726 & $<$4   & $<$0.2 & K$_{s}$=12.6, $P = 5850$~s$^{\mathrm{c}}$ & \cite{revnivtsev03}\\
IGR J16393--4643 & 16.8   & 0.54   & AX J1639.0--4642 (3EG?) & \cite{malizia04}\\
IGR J16418--4532 & 16.8   & 0.54   & -- & \cite{tomsick04}\\
IGR J16479--4514 & 11.6   & 0.39   & -- & \cite{molkov03}\\
IGR J16558--5203 & 4.3    & 0.17   & -- & \cite{walter04}\\
IGR J17091--3624 & $<$4   & $<$0.3 & 1SAX J1709--36, radio & \cite{kuulkers03}\\
IGR J17195--4100 & 7.1    & 0.36   & -- & \cite{walter04}\\
IGR J17252--3616 & 10.8   & 0.82   & -- & \cite{walter04}\\ 
\hline \\
\end{tabular}
\end{center}
\begin{list}{}{}
\footnotesize
\vspace{-0.7cm}
\item[$^{\mathrm{a}}$] Here, we give the ISGRI count rate in the 
20-40 keV band.  The {\em RXTE} and {\em INTEGRAL} observations of 
4U 1630--47 reported below can be used to obtain an approximate flux 
calibration.  For 4U 1630--47, the 20-40 keV flux measured by HEXTE 
is $2.7\times 10^{-10}$ erg~cm$^{-2}$~s$^{-1}$, and the ISGRI count 
rate is 3.75 c/s.  Thus, 1.0 ISGRI c/s $\sim$ $7.2\times 10^{-11}$ 
erg~cm$^{-2}$~s$^{-1}$ (20-40 keV) $\sim$ 6.5 mCrab.  
\vspace{-0.2cm}
\item[$^{\mathrm{b}}$] See Chaty \& Filliatre in these proceedings.
\vspace{-0.2cm}
\item[$^{\mathrm{c}}$] Pulsations reported in \cite{patel04}.
\end{list}
\end{table*}

\subsection{4U 1630--47}

\begin{figure}[ht]
\centerline{\includegraphics[width=1.0\linewidth,angle=0]{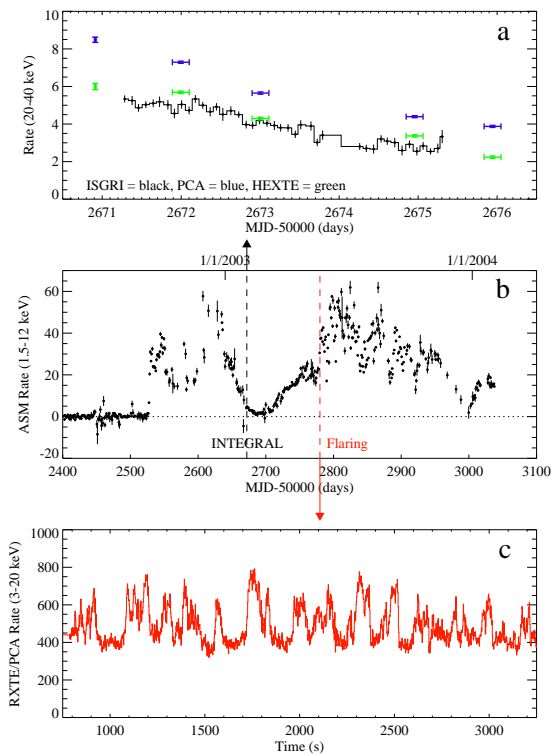}}
\caption{4U 1630--47 X-ray light curves measured by (a) 
ISGRI, PCA, and HEXTE, (b) the RXTE ASM, and (c) the PCA. 
The ASM light curve for the current outburst is shown in (b).  
20-40 keV light curves are shown in (a).  An example of the 
4U 1630--47 X-ray flares is shown in (c).\label{fig:lc3}}
\end{figure}

Figure~\ref{fig:lc3}b shows the {\em RXTE} All-Sky Monitor
(ASM) 1.5-12 keV light curve for 4U 1630--47 for the current 
outburst (which is still in progress as of 2004 March).  
The peak flux of about 60 ASM is about 50\% higher than
the peak of the previous four outbursts occurring during
the {\em RXTE} era.  The 2003 February {\em INTEGRAL}
observation (marked on Figure~\ref{fig:lc3}b) occurred
during a time when the soft X-ray flux was decaying, but
the hard X-ray flux was still relatively strong.  
Although strong, Figure~\ref{fig:lc3}a shows that the 
20-40 keV flux declined during the observation.  The
nearly continuous ISGRI light curve indicates that the
decline proceeded in a relatively gradual manner, and
this figure also shows the 20-40 keV count rates for
the PCA and HEXTE for the five {\em RXTE} pointings.

\begin{figure}[ht]
\centerline{\includegraphics[width=1.0\linewidth,angle=0]{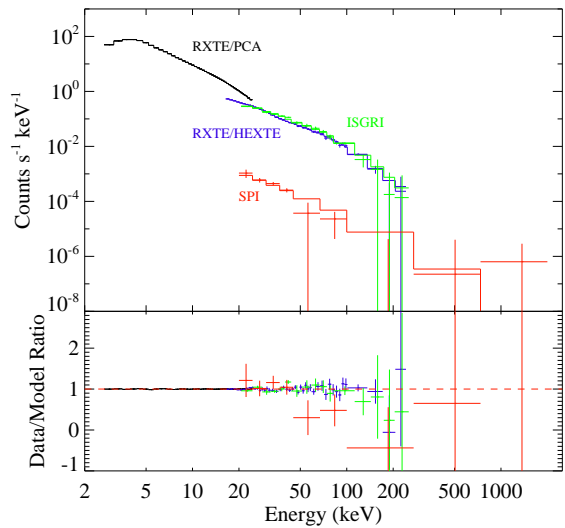}}
\vspace{-1.25cm}
\caption{The INTEGRAL and RXTE 3-2000 keV energy spectrum.  The 
exposure time for INTEGRAL is 293 ks, while the exposure time from 
the PCA is 43 ks, and the exposure time for HEXTE is 28 ks.  
\label{fig:spectrum}}
\end{figure}

We extracted energy spectra for the ISGRI and SPI
instruments on-board {\em INTEGRAL} and for the 
{\em RXTE} instruments PCA and HEXTE, and the fitted
spectrum is shown in Figure~\ref{fig:spectrum}.  
The spectrum is fitted with a model that has been
used for previous outbursts \citep[e.g.,][]{tk00}
consisting of a disk-blackbody with $kT_{in} = 
0.80\pm 0.03$ keV plus a power-law with a photon 
index of $\Gamma = 2.09\pm 0.03$.  The model also 
includes interstellar absorption, a smeared iron 
edge, an iron emission line, and a high energy 
cutoff with a cutoff energy of $27\pm 4$ keV and 
an exponential folding energy of $233\pm 40$ keV.
The PCA normalization is fixed to 1.0 and the other 
instrument normalization are left free, giving 
$0.811\pm 0.005$ for HEXTE, $0.60\pm 0.01$ for ISGRI, 
and $1.36\pm 0.13$ for SPI.  Thus, the normalizations
are not consistent between instruments.  We have 
extracted power spectra using the PCA data.  The 
power spectrum is well-described by a band-limited
noise component plus a power-law at low frequencies, 
and the 0.1-10 Hz fractional rms amplitude is 4\%.
The moderate level of timing noise and the strong 
power-law component in the energy spectrum 
indicate that the source was in an intermediate
state between the canonical Soft and Hard states
\citep[See][for a recent review of black hole
spectral states.]{mr03}.

For the five {\em RXTE} observations, $kT_{in}$ 
and $\Gamma$ range from 0.74-0.82 keV and 2.11-2.18, 
respectively, but neither parameter correlates with 
the overall flux.  Also, it is notable that the
total disk-blackbody flux does not correlate with 
the total flux.  The change in the total flux is 
driven by the change in the power-law normalization.  
The other quantity that is significantly different
between observations is the high energy cutoff.  A 
cutoff is only required at high significance for the 
second spectrum from the top, and the cutoff is not 
as sharp for the bottom two spectra.  Often BHCs pass 
quickly through intermediate states, and these 
observations represent a useful study of the spectral 
evolution during intermediate states.

For much of the current 2002-2004 outburst, 4U 1630--47
has shown dramatic X-ray flaring behavior that is uncommon, 
although rapid, high amplitude flux changes have been 
reported for at least one previous outburst \citep{dieters00}.  
Figure~\ref{fig:lc3}c shows an example of this behavior
from a 2003 May {\em RXTE} observation.  Although not
shown in this work, we made a hardness-intensity diagram
for this light curve, indicating that there is a tight
correlation between hardness and intensity. 

\begin{figure}[ht]
\centerline{\includegraphics[width=1.0\linewidth,angle=0]{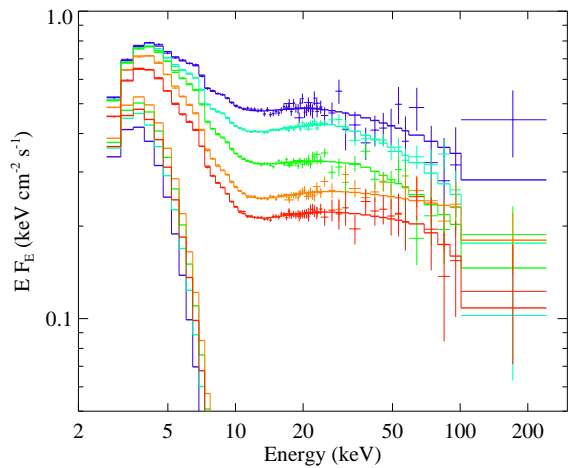}}
\caption{RXTE PCA and HEXTE energy spectra for the five observations
made during or close to the time of the INTEGRAL observation.
Figure~\ref{fig:lc3}a shows that the flux was gradually falling, and
the spectra are in chronological order from top to bottom.
\label{fig:evolution}}
\end{figure}

\vspace{1.0cm}
\section{Summary and Conclusions}

\noindent
$\bullet$~Fifteen of the 38 IGR sources are in the 4U 1630--47 field.
Only three of the sources are not detected, indicating that as many
as 12 of the sources are persistent.  It is possible that a subset
of the IGR sources are part of a new population of persistent, hard, 
intermediate luminosity X-ray sources.\\
$\bullet$~For much of its current outburst, 4U 1630--47 has shown
dramatic flaring that is reminiscent of the flaring seen from
GRS 1915+105.  For 4U 1630--47, the hardness and intensity are
correlated during the flares, indicating that these are hard
flares that are related accretion (and possibly ejection) 
mechanisms rather than being due to, e.g., absoption from the
outer edge of the disk.\\
$\bullet$~The ISGRI light curve (Figure~\ref{fig:lc3}a) and the
spectra shown in Figure~\ref{fig:evolution} show that 4U 1630--47
spectrum gradually changed, and the main change is in the 
normalization of the power-law component.\\
$\bullet$~The {\em RXTE} spectra indicate that the high energy
cutoff also changes during the observations, and this sort of
evolution must be kept in mind when combining {\em INTEGRAL}
data from long observations.

\section*{Acknowledgments}

JAT would like to thank Jerome Rodriguez, Luigi Foschini, and
Katja Pottschmidt for help with the {\em INTEGRAL} data analysis.
We also would like to thank Arvind Parmar for his help in getting
this {\em INTEGRAL} Target of Opportunity observation done.  JAT
acknowledges partial support from NASA grant NAG5-12703.
This research has made use of the SIMBAD database, operated at 
CDS, Strasbourg, France


\end{document}